# Compensation Behavior and Magnetic Properties of a Core/Shell Double Fullerene Structure

S. Aouini [1], A. Mhirech [1], A. Alaoui-Ismaili [2] and L. Bahmad [1,*]

[1] Laboratoire de Matière Condensée et Sciences Interdisciplinaires (LaMCScI)

Mohammed V University of Rabat, Faculty of Sciences, B.P. 1014, Rabat, Morocco

[2] Département de Mathématiques, Mohammed V, University of Rabat, Faculty of Sciences, B.P. 1014, Rabat, Morocco

**Abstract:**

In This work, we apply the Monte Carlo simulations to study the magnetic properties and compensation behavior of a core/shell double fullerene structure $X_{60}$ where the symbol X can be assigned to any magnetic atom. We focus our study on a system formed by a double sphere forming a core-shell. The two spheres are containing the spins: $S = 1/2$ in the core; and $\sigma = 1$ in the shell, respectively. In a first step, we investigate and discuss the corresponding ground state phase diagrams in different planes.

Also, we illustrated the behavior of the magnetizations and the effect of the coupling exchange interactions as well as the crystal field. The effect of the external magnetic field, and the exchange coupling interactions on the hysteresis loops have been inspected. To complete this study, we showed the existence of the compensation temperature for the studied system.

[*] Corresponding author: bahmad@fsr.ac.ma; lahou2002@gmail.com.

## 1- Introduction:

Recently, the fullerenes have been discovered for the first time. These nano-structures are formed principally by carbon atoms. Several fullerenes were discovered containing more and fewer carbon atoms. Among them, the Buckyball, containing 60 carbon atoms, is the most popular fullerene $C_{60}$. In fact, the fullerene $C_{60}$ remains not only the easiest one to produce, but also the cheapest and commercialized one. The structure of the "$C_{60}$" is a combination of 12 pentagonal and 20 hexagonal rings, forming a spheroid shape with 60 vertices containing 60 carbons [1]. The structure and the production of the fullerenes, as well as their chemical, mechanical, magnetic and optical properties make them outstanding molecules in several applications, particularly in nanotechnology sciences [2, 3]. The various applications of fullerenes depend on their particular and specific properties [4, 5]. These nano-structures have also exceptional properties with a very long electron spin relaxation times [6]. This makes them essential components in the future for nano-electromechanical systems. The fullerenes are essential for the engineering flexibility and promising applications in electrical, military and medical fields [7, 8].

Also, the fullerenes have become important compounds in science, nanotechnology and in biological activities. Nowadays, for their practical properties, the fullerenes are very useful in many interesting scientific fields [9-14]. In fact, several outstanding works focus on these nano-structures in order to study and interpret their magnetic properties [15-20].

On the other hand, several applications including computer simulations [21, 22], have been developed in this field of research. Theoretically, different studies have been focused on these nanostructures, to show their magnetic and/or ferromagnetic properties, using different methods such as the effective field theory [23-25], the mean field [26] and the Monte Carlo simulations [27-30]. The use of Monte Carlo simulations showed many characteristic phenomena occurring in such systems.

Some recent studies [31-35] have focused on these nano-structure systems, showing their rich critical behavior and many fascinating phenomena such as the magnetic wetting and the layering transitions. In the last decade, much works have been focused on the study of other nano-structures including alternate layers [36-41]. The obtained results are very interesting when performed on the critical behavior in these nano-particles. In particular, interesting magnetic phenomena have been found, especially in the region near the compensation temperature $T_{comp}$.

The aim of this work is to apply the Monte Carlo method to investigate the magnetic properties and the corresponding phase diagrams. This study will complete the earlier studies, for other different systems, concerning the behavior of the magnetizations [42] and the hysteresis cycles [43]. Our calculations are based on the influence of the magnetic field and the coupling exchange interactions and their impact on the magnetic properties the double fullerene $C_{60}$. In this work, we have examined the effect of the crystal field on such nano-structures. This paper is organized as follows: in section 2, we present the model and the theoretical formulations. In section 3, we discuss the obtained numerical results. Finally, section 4 draws the conclusions.

## 2- Model and method:

In this part, we give the formulations used in the study of the magnetic properties of the double fullerene structure $X_{60}$. The symbol "X" can be assigned to any magnetic atom. In Fig. 1, we present a scheme of the studied system. The corresponding geometry is formed by two spherical layers consisting of two types of spins. The core is formed by the spins $S = 1/2$, whereas the shell is consisting of the spins $\sigma = 1$.

The Hamiltonian governing the studied system is defined by:

$$\mathcal{H} = -J_{sh} \sum_{i,j} \sigma_i \sigma_j - J_c \sum_{k,l} S_k S_l - J_{int} \sum_{i,k} \sigma_i S_k - H \sum_i (\sigma_i + S_i) - D \sum_i \sigma_i^2 \qquad (1)$$

where, $J_{sh}$ and $J_c$ are the exchange couplings between the first nearest neighbor atoms with spin $\sigma - \sigma$ and $S - S$, respectively. The exchange coupling interaction $J_{int}$ is related to the nearest neighbor magnetic atoms with spin $S$ and $\sigma$, belonging to the shell and the core, respectively. D is the crystal field applied only on the spins $\sigma$. The parameter H stands for the external magnetic field.

In this study, we apply the Monte Carlo method under the Metropolis algorithm [44-48]. A number of $10^6$ Monte Carlo steps per spin are performed, generating new configurations according to the Boltzmann distribution. We discard the first $10^5$ Monte Carlo steps, and average over different initial conditions. The error bars were calculated with a Jackknife method [49, 50].

Hereafter, we define the studied physical parameters as follows:

The magnetizations per spin are:

$$m_s = \frac{1}{N_s}\sum_{i=1}^{N_s} S_i \qquad \text{in the core} \qquad (2)$$

$$m_\sigma = \frac{1}{N_\sigma}\sum_{i=1}^{N_\sigma} \sigma_i \qquad \text{in the shell} \qquad (3)$$

With: $N_s = 21$ and $N_\sigma = 39$ spins, respectively.

At the compensation point, the condition $|m_s(T_{comp})| = |m_\sigma(T_{comp})|$ must be verified.

The total magnetization is defined as:

$$M = \frac{N_\sigma \cdot m_\sigma + N_s \cdot m_s}{N_T} \qquad (4)$$

Where the total number of spins is:

$$N_T = N_s + N_\sigma = 60 \qquad (5)$$

The total magnetic susceptibility $\chi$ is defined as:

$$\chi = \beta N_T (<M^2> - <M>^2) \qquad (6)$$

Where: $\beta = \frac{1}{k_B T}$, with T being the absolute temperature and ($k_B = 1$) is the Boltzmann's constant.

The internal energy per site is given by:

$$E = \frac{1}{N}\langle \mathcal{H} \rangle \qquad (7)$$

3- **Results and discussion:**

3-1- **Ground state phase diagrams**

In this section, we study the magnetic properties of the studied system, in the absence of any temperature ($T = 0\ K$). In fact, the ground state phase diagrams are calculated relying on the parameters defined by the Hamiltonian of Eq. (1). Indeed, the all possible energy configurations $2 \times 3 = 6$ are computed and compared in the corresponding ground phase diagrams, plotted in Figs. 2(a)-(d).

A perfect symmetry is found in the plane (D, H) for the fixed values of the exchange coupling interactions: $J_c = 1$, $J_{sh} = 0.1$ and $J_{int} = -0.1$. The all possible phases are found for the negative values of the crystal field D whereas only two phases are found to be stable for the positive values of this parameter with large regions, see Fig. 2a.

In the absence of the external magnetic field (H = 0), some new phases are appearing in the plane $(D, J_c)$ for the fixed values of the exchange coupling interactions: $J_{sh} = 0.1$ and $J_{int} = -0.1$, see Fig. 2b, the all phases are found for the negative values of the crystal field, where only two phases are found to be stable for the positive values of this parameter.

In Fig. 2c, we found in the plane $(D, J_{int})$ the all possible configurations for the negative values of the crystal field, where only four phases are found to be stable for the positive values of this parameter with large regions.

In the plane $(D, J_{sh})$ for the fixed values of the exchange coupling interactions: $J_c = 1$ and $J_{int} = -0.1$. A perfect symmetry is found, regarding the crystal field D in the absence of the external magnetic field for the fixed values of the exchange coupling interactions: $J_{sh} = 0.1$ and $J_c = 1$, and in the absence of the external magnetic field (H=0), we found the same phases showed in fig. 2b, where two phases are found to be stable for the positive values of the crystal field and two phases are found to be stable for the negative values of this parameter with large regions, see Fig. 2d.

### 3-2- Monte Carlo Study

In this part, we use Monte Carlo simulations for non-null values of temperature $(T \neq 0)$ to investigate the effect of the exchange coupling interactions, the external magnetic and the crystal fields. We will study the effect of increasing the exchange coupling interactions $J_{int}$ at which the total magnetization as a function of the external magnetic field H. In fact, the corresponding phase diagrams are plotted in Figs. 3(a)-(d) and Figs. 4(a)-(d).

In Figs. 3(a)-(d), for the fixed value of the temperature (T = 0.5), and for the fixed values of the exchange coupling interactions: $J_{sh} = 0.1$ and $J_c = 1$, we show the dependence of the hysteresis cycles in the absence of the crystal field (D = 0), at the value of the exchange coupling interactions ($J_{int} = -0.05$), see Fig. 3(a), we found a small surface of the external magnetic field, when the studied system is named a soft system. The behavior of the total magnetization of each layer is of second order transition type from the negative values to the positive values of the total magnetization. By increasing the value of the exchange coupling interactions ($J_{int} = -0.1$), the behavior of the total magnetization of each layer is of second order transition type from the negative values to the positive values of the total magnetization.

The surface of the external magnetic field becomes biggest, but the studied system is named a soft system, see Fig. 3(b). By increasing the value of the exchange coupling interactions ($J_{int} = -0.2$), see Fig. 3(c), we found a large surface of the external magnetic field, when the studied system is named a hard system. The behavior of the total magnetization of each layer is of first order transition type from the negative values to the positive values of the total magnetization. By increasing the value of the exchange coupling interactions ($J_{int} = -0.4$), the behavior of the total magnetization of each layer is of first order transition type from the negative values to the positive values of the total magnetization, but the surface of the external magnetic field becomes very large, when the studied system is named a hard system, see Fig. 3(d).

In Figs. 4(a)-(d), for the fixed value of the temperature (T = 0.5), and for the fixed values of the exchange coupling interactions: $J_{sh} = 0.1$ and $J_c = 1$, we show the dependence of the hysteresis cycles in the presence of the crystal field (D = 1), at the value of the exchange coupling interactions ($J_{int} = -0.05$), see Fig. 4(a), the behavior of the total magnetization of each layer is of second order transition type from the negative values to the positive values of the total magnetization. The surface of the external magnetic field becomes biggest, but the studied system is named a soft system. By increasing the value of the exchange coupling interactions ($J_{int} = -0.1$), we found a small surface of the external magnetic field, when the studied system is named a soft system. The behavior of the total magnetization of each layer is of second order transition type from the negative values to the positive values of the total magnetization. See Fig. 4(b). By increasing the value of the exchange coupling interactions ($J_{int} = -0.2$), see Fig. 4(c), the behavior of the total magnetization of each layer is of first order transition type from the negative values to the positive values of the total magnetization, but the surface of the external magnetic field becomes large, when the studied system is named a hard system. By increasing the value of the exchange coupling interactions ($J_{int} = -0.4$), we found a very large surface of the external magnetic field, when the studied system is named a hard system. The behavior of the total magnetization of each layer is of first order transition type from the negative values to the positive values of the total magnetization. See Fig. 4(d).

To investigate the effect of varying the exchange coupling interactions ($J_{int} = -0.1, -0.5, -0.7, -1$) on the behavior of the studied system, as a function of the crystal field $D$, we plot in Fig. 5, the corresponding magnetization profiles. This figure is plotted for a fixed temperature value (T = 0.3), and also for fixed values of the exchange coupling interactions: $J_{sh}$ = 0.1 and $J_c$ = 1. The sphere layers of the studied system transit simultaneously and a first order transition type is found. Indeed, the magnetization profiles transit from negative values to positive values, in the absence of the external magnetic field (H = 0). This behavior can be explained by the low temperature effect. In fact, it is well known that all the transitions of such systems are of first order type at very low temperature values. Hence, the crystal field effect is only to change simultaneously the orientation and sign of the global magnetizations.

In this work, we are especially interested in the effect of the exchange interaction $J_{int}$ on the compensation behavior, for the studied system, appearing for specific values of the temperature. The effect the other physical parameters, such as the crystal and external magnetic fields, is illustrated in some of our previous works, see, Refs. [31-35].

We will also examine the effect of increasing the temperature at which the total magnetization as a function of the exchange coupling interactions $J_{int}$. In fact, the corresponding phase diagrams of the magnetizations are plotted in Fig. 6. In the absence of both of the external magnetic and crystal fields (H = 0 and D = 0), for the fixed values of the exchange coupling interactions: $J_{sh}$ = 0.1 and $J_c$ = 1. The behavior of the total magnetization is of first order transition type from the negative values to the positive values of the total magnetization for ($T/J_c$ = 0.2, 0.4), while, this behavior is of second order transition type from the negative values to the positive values of the total magnetization for ($T/J_c$ = 0.5, 0.6, 0.65).

We are interested in the compensation behavior of our system. For this reason, we inspect the existence of the compensation temperature $T_{comp}$ for the fixed values of the exchange coupling interactions: $J_c$ = 1, $J_{sh}$ = 0.1 and $J_{int}$ = −0.1 in the absence of the external magnetic and the crystal fields (H = 0 and D = 0). In fact, we plot in Fig. 7 the corresponding phase diagrams of the magnetizations as a function of the temperature. We found in this figure just one compensation temperature ($T_{comp} \cong 0.5$) depending on the magnetizations of each layer and the total magnetization absolute values. These parameters annul and stabilize for T ≥ 0.75. While, we plot in Fig. 8 the temperature as a function of the exchange coupling interactions $J_{int}$ for the fixed values of the exchange coupling interactions: $J_c$ = 1, $J_{sh}$ = 0.1 in the absence of the external magnetic and the crystal fields (H = 0 and D = 0). We both the compensation and the critical temperatures. When increasing the exchange coupling interactions absolute values the compensation behavior of the studied system is strongly governed by the effect of the parameter $J_{int}$.

Furthermore, it is well known that the parameter $J_{int}$, representing the interaction between the ferri-magnetic spin moments belonging to the shell and the core of the spheres, is the main physical parameter responsible on the compensation behavior. This phenomenon is appearing only for specific values of the temperature, less than the critical temperature. This is called the super-paramagnetic behavior. This situation disappears rapidly for increasing temperature values giving rise to the standard Curie temperature. The compensation and Curie temperatures are found to be confused for large absolute values of the parameter $J_{int}$, see Fig. 8. In other words, the compensation behavior, is only found for weak coupling values of the exchange interactions between the spin moments of the core and the shell.

4- **Conclusion:**

In this work, we have investigated the magnetic properties in a double fullerene structure based on two spheres consisting of core-shell. The effect of the external magnetic, the crystal fields and the coupling exchange interactions is also illustrated and discussed in this paper. The ground state phase diagrams in different planes were presented for different values of the phase space parameters. In the absence of the external magnetic field, we found that the most stable configurations are present for negative values of the crystal filed. On the other hand, in the presence of the crystal field it is found that a perfect symmetry appears regarding the external magnetic field H. Also, the behavior of the magnetizations and the hysteresis cycles have been analyzed in several phase diagrams by using the Monte Carlo simulations. We have showed the existence of the compensation and critical temperatures of the studied system. The hysteresis loops showed that when increasing the temperature values the behavior of the total magnetization changed from the first to the second transition type. The compensation temperature values depends strongly on the exchange coupling interactions for this nano-structure. This interesting result is illustrated in different phase diagrams.

# Fig. 1

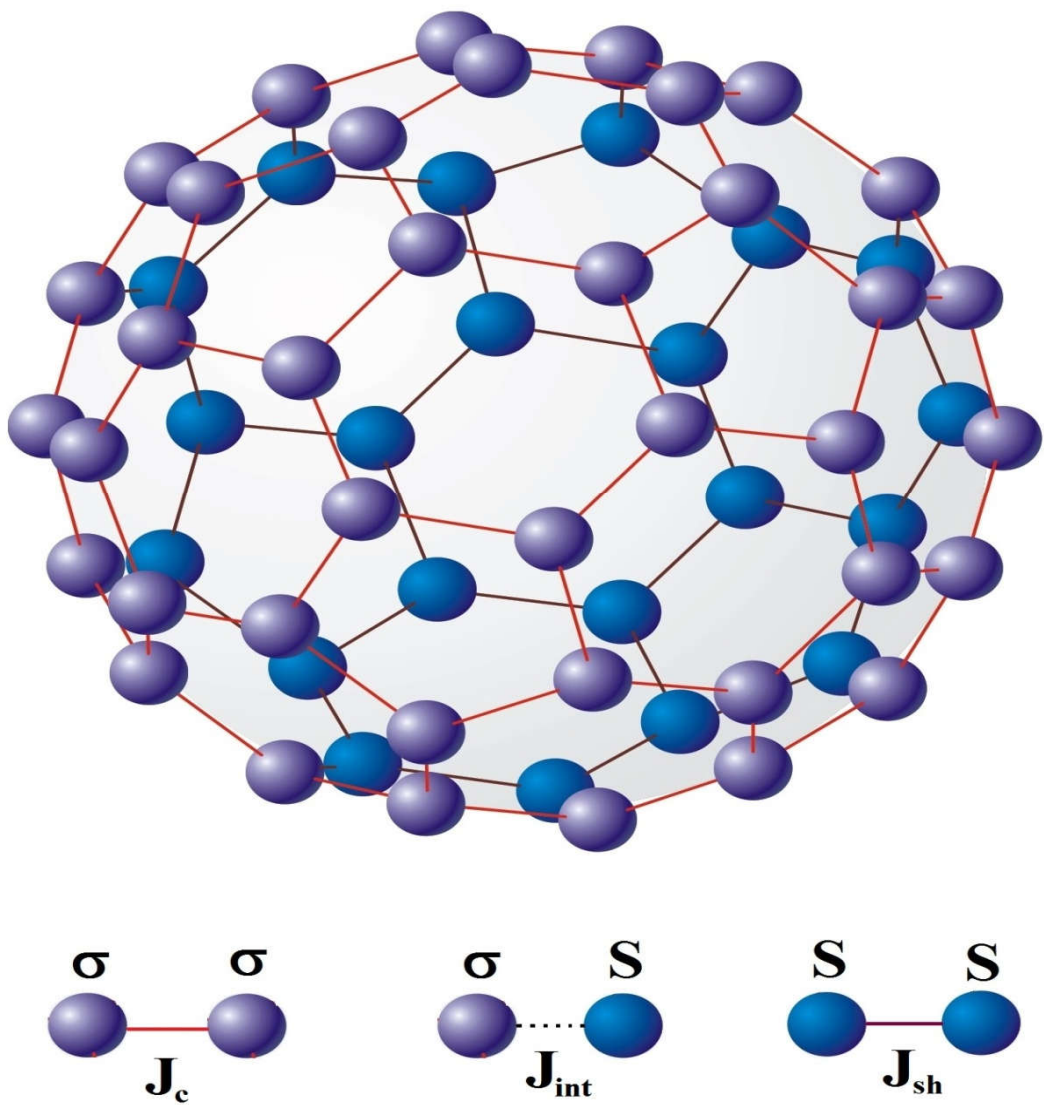

**Fig. 1:** Geometry of the studied system formed by two sphere layers with spins $\sigma = 1$ and $S = 1/2$, containing: $N_S = 21$ and $N_\sigma = 39$, and $N_T = N_S + N_\sigma = 60$.

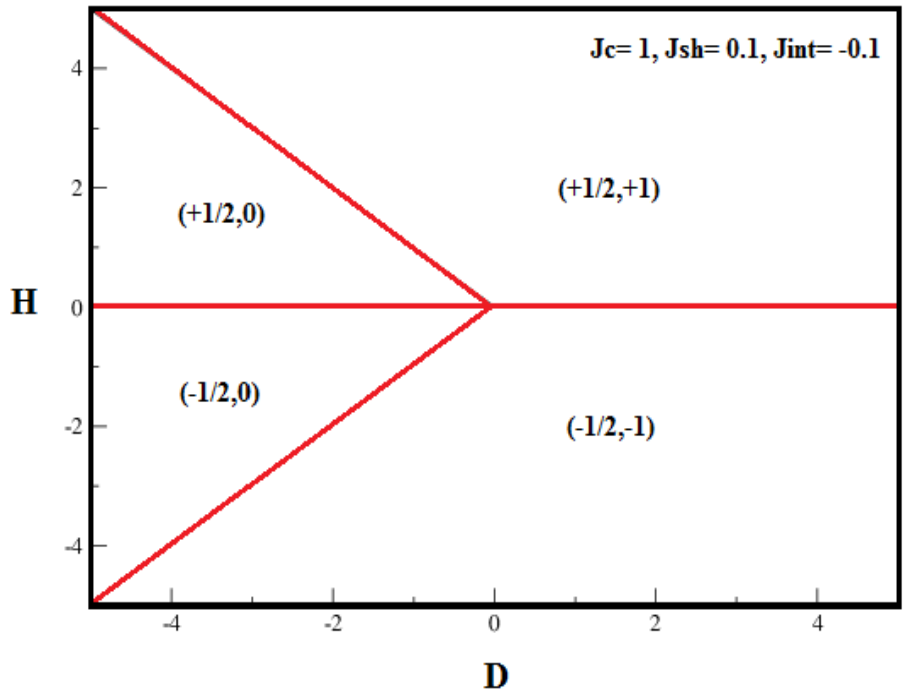

**Fig. 2:** Ground state phase diagrams of the studied system in the plane (D, H) for $J_c = 1$, $J_{sh} = 0.1$ and $J_{int} = -0.1$.

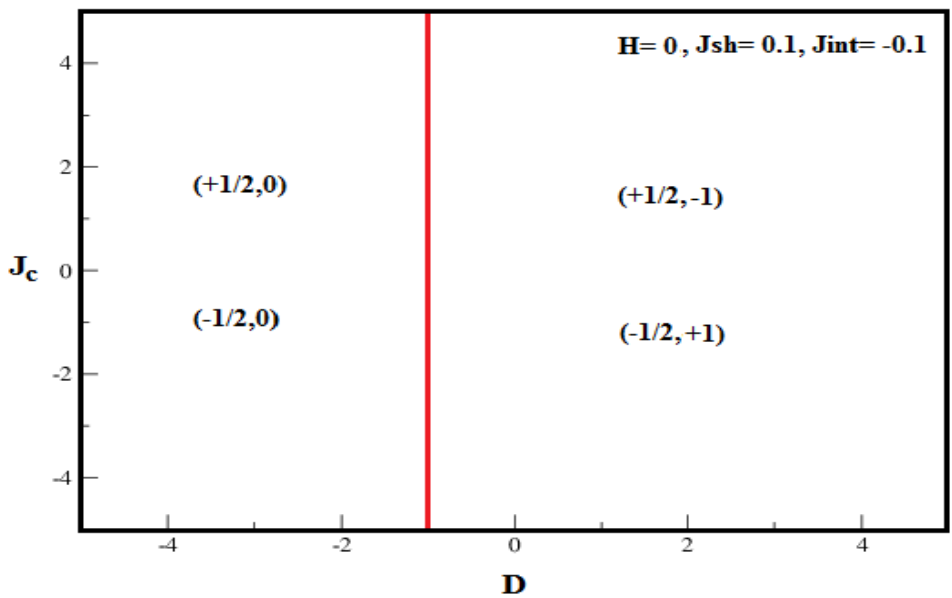

**Fig. 2:** Ground state phase diagrams of the studied system in the plane (D, $J_c$) for $H = 0$, $J_{sh} = 0.1$ and $J_{int} = -0.1$.

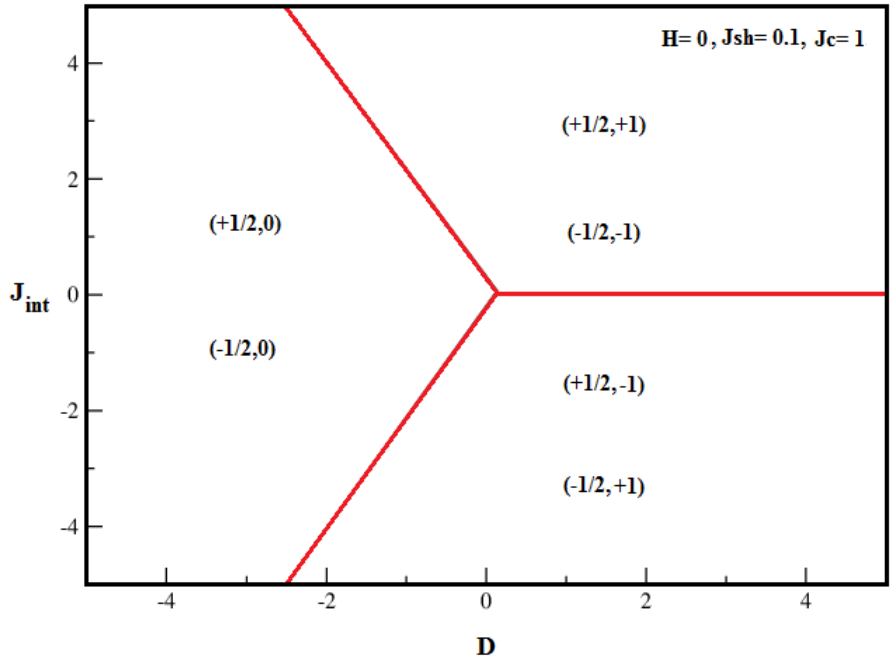

**Fig. 2:** Ground state phase diagrams of the studied system in the plane $(D, J_{int})$ for $H = 0$, $J_{sh} = 0.1$ and $J_c = 1$.

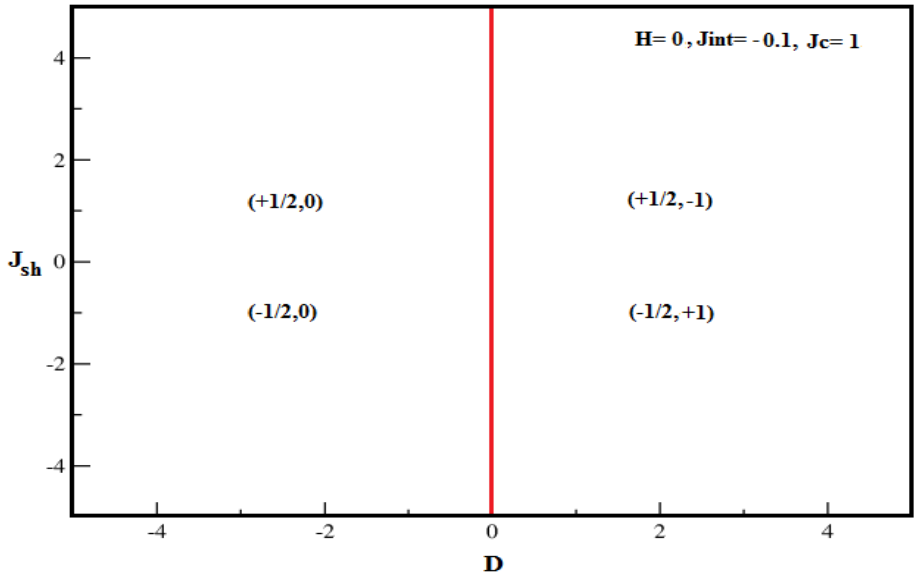

**Fig. 2:** Ground state phase diagrams of the studied system in the plane $(D, J_{sh})$ for $H = 0$, $J_{int} = -0.1$ and $J_c = 1$.

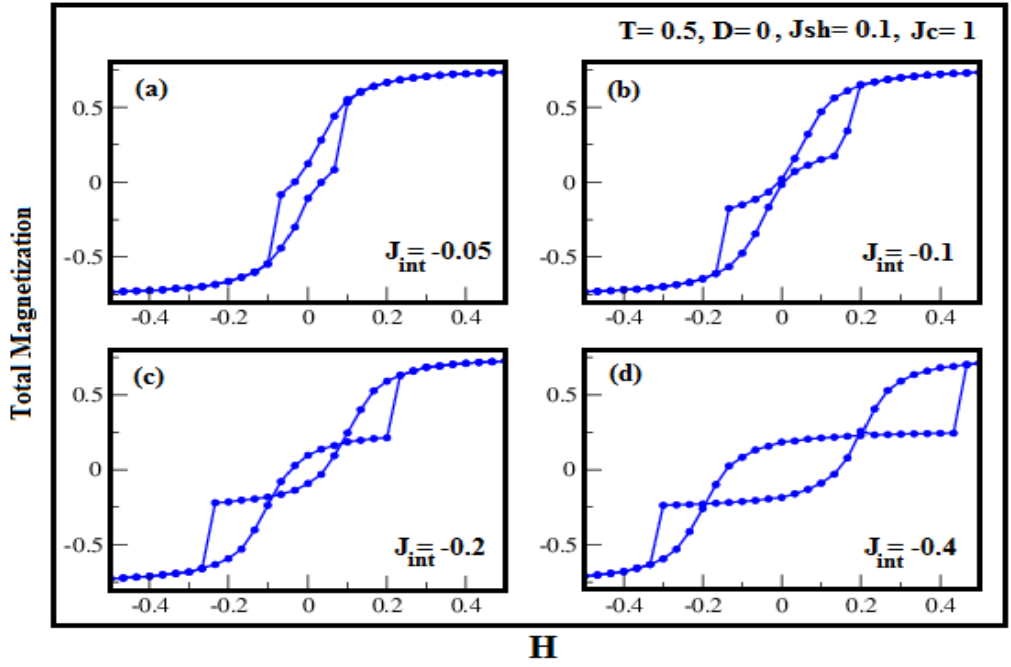

**Fig. 3:** The total magnetization profiles as a function of H, with: $J_{sh} = 0.1$, $J_c = 1$ and $D = 0$, $T = 0.5$ for: (a): $J_{int} = -0.05$, (b): $J_{int} = -0.1$, (c): $J_{int} = -0.2$, and (d): $J_{int} = -0.4$.

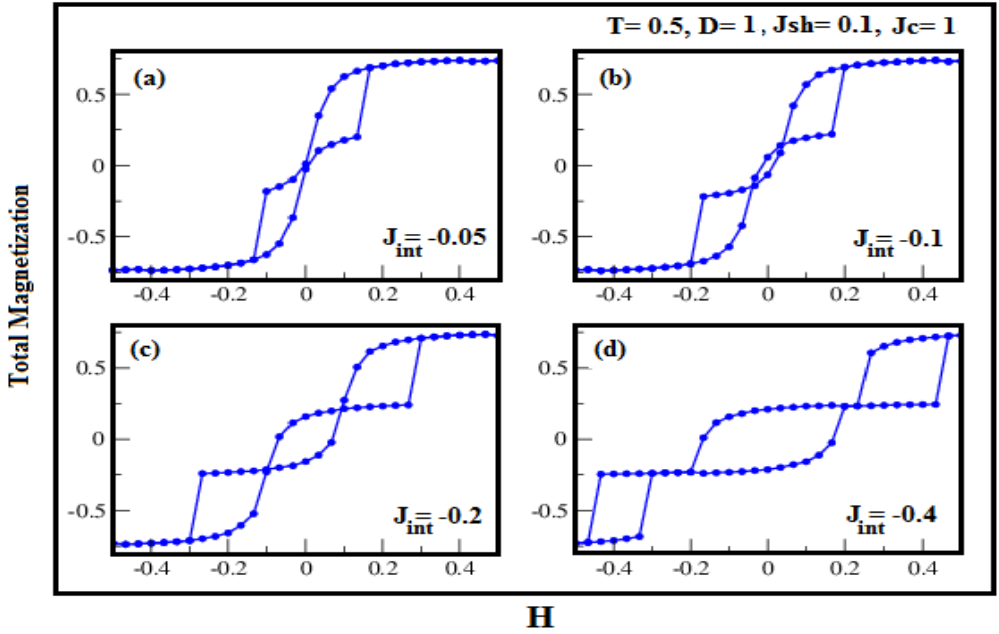

**Fig. 4:** The total magnetization profiles as a function of H, with: $J_{sh} = 0.1$, $J_c = 1$ and $D = 1$, $T = 0.5$ for: (a): $J_{int} = -0.05$, (b): $J_{int} = -0.1$, (c): $J_{int} = -0.2$, and (d): $J_{int} = -0.4$.

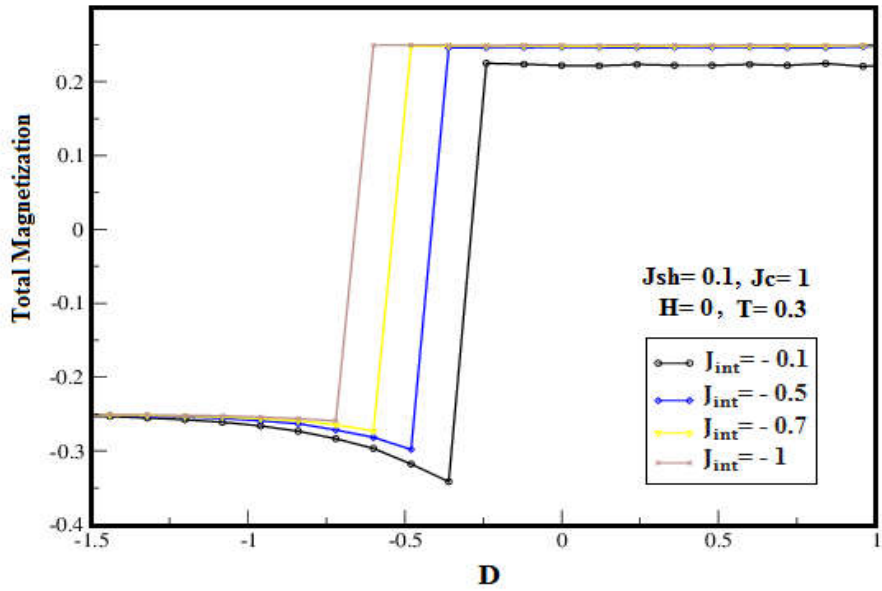

**Fig. 5:** The total magnetization profiles as a function of the crystal field D, with: $J_{sh} = 0.1$, $J_c = 1$ and $H = 0$, $T = 0.3$ for $J_{int} = -0.1$, $J_{int} = -0.5$, $J_{int} = -0.7$, and $J_{int} = -1$.

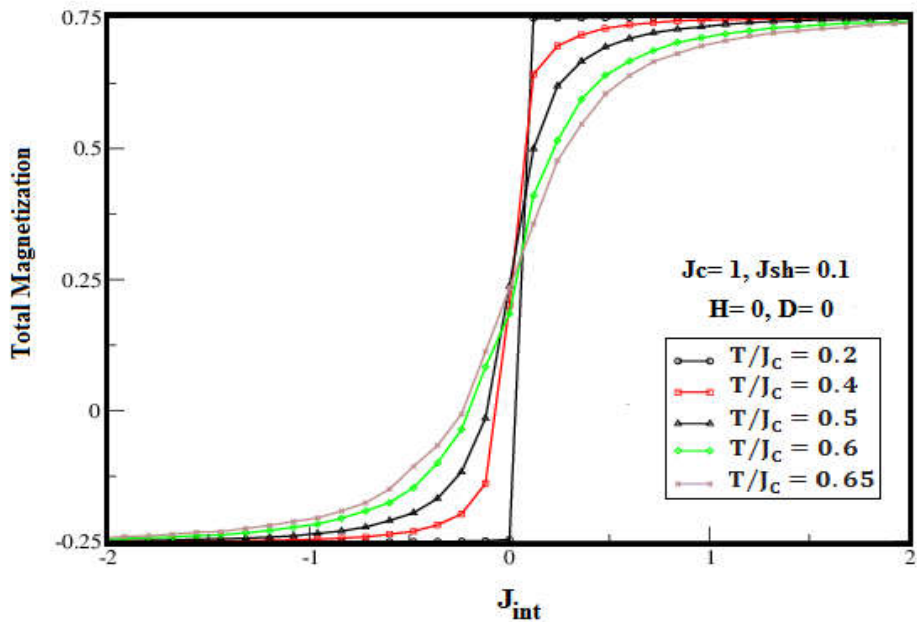

**Fig. 6:** The total magnetization profiles as a function of $J_{int}$, with: $J_c = 1$, $J_{sh} = 0.1$ and $H = 0$ and $D = 0$ for $T/J_c = 0.2$, $T/J_c = 0.4$, $T/J_c = 0.5$, $T/J_c = 0.6$ and $T/J_c = 0.65$.

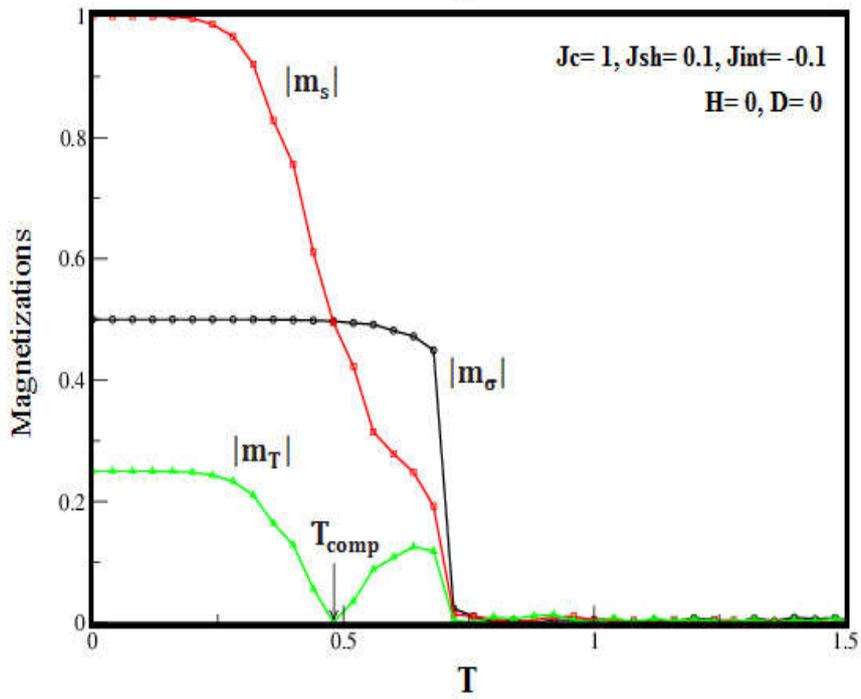

**Fig. 7:** The magnetizations profiles as a function of T, with: $J_c = 1$, $J_{sh} = 0.1$ and $J_{int} = -0.1$ for $D = H = 0$.

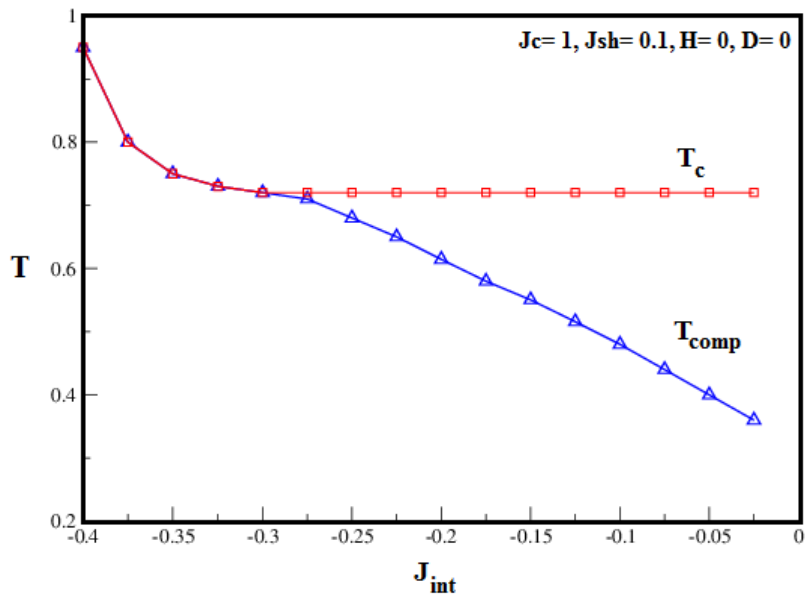

**Fig. 8:** Temperature profiles in terms of $J_{int}$, for $J_c = 1$, $J_{sh} = 0.1$ and $H = 0$ and $D = 0$.